\documentclass{article}

\usepackage[english]{babel}

\usepackage[letterpaper,top=2cm,bottom=2cm,left=3cm,right=3cm,marginparwidth=1.75cm]{geometry}

\usepackage{amsmath}
\usepackage{graphicx}
\usepackage{booktabs}       % professional-quality tables
\usepackage{siunitx}
\usepackage{setspace} % 用于调整行间距
\usepackage{caption}   % 用于调整caption与表格之间的距离
\usepackage{float}
\usepackage{multirow}
\usepackage{array}
\usepackage{pythonhighlight}
\usepackage[colorlinks=true, allcolors=blue]{hyperref}

\title{Memory Analysis on the Training Course of DeepSeek Models}
\author{
Ping Zhang, Lei Su \\
\\
\small Baichuan-Inc \\
\small \texttt{zhangping@baichuan-inc.com} \\
}
\date{}

\begin{document}
\maketitle

\begin{abstract}
We present a theoretical analysis of GPU memory consumption during the training of DeepSeek models such as DeepSeek-v2 and DeepSeek-v3. Our primary objective is to clarify the device-level memory requirements associated with various distributed training configurations. Specifically, we examine critical factors influencing memory usage, including micro-batch size, activation recomputation policies, 3D parallelism, and ZeRO optimizations. It is important to emphasize that the training policies discussed in this report are not representative of DeepSeek's official configurations. Instead, they are explored to provide a deeper understanding of memory dynamics in training of large-scale mixture-of-experts model.
\end{abstract}

\section{Overview Architecture}
\subsection{Transformer Block}
This report primarily focuses on analyzing the structural details of DeepSeek-v3 \cite{liu2024deepseekv3}, which currently represents the state-of-the-art among open-source models. While the report addresses DeepSeek-v3, it is equally applicable to DeepSeek-v2 \cite{liu2024deepseekv2} and can also serve as a reference framework for analyzing general mixture-of-experts (MoE) models.

\begin{figure}[!h]
    \centering
    \includegraphics[width=0.99\linewidth]{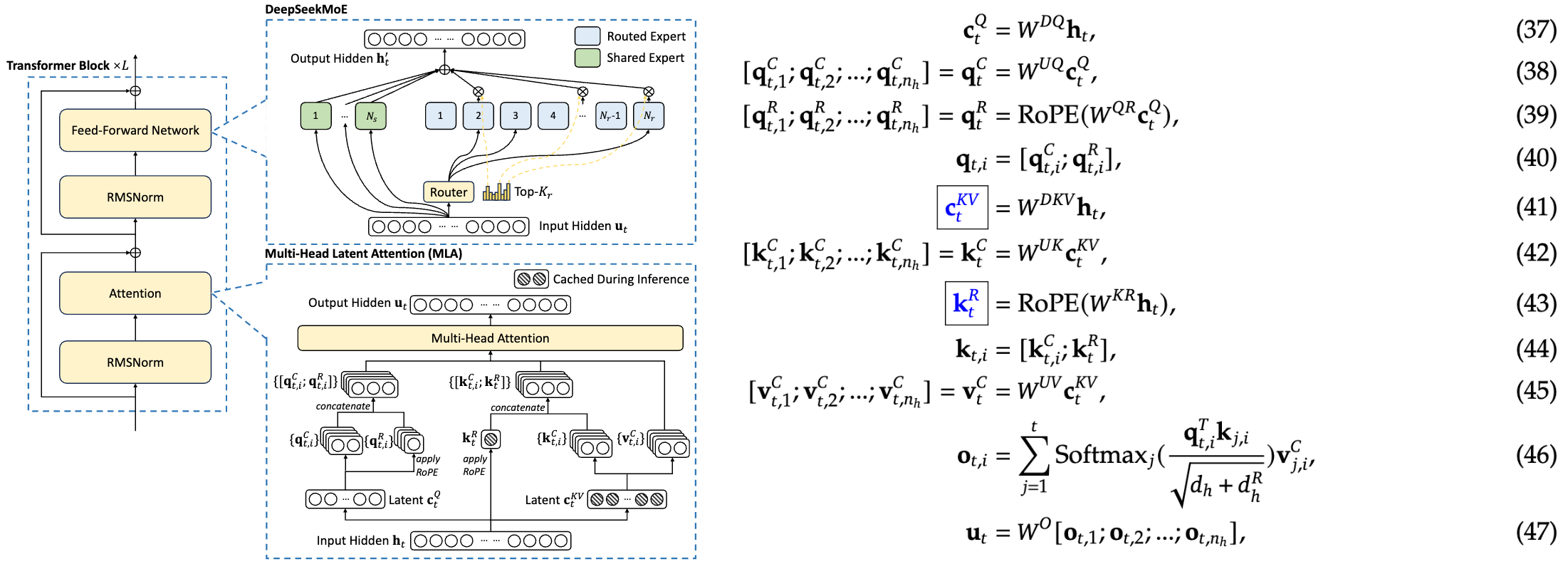}
    \caption{Illustration of the basic architecture of DeepSeek-v3 \cite{liu2024deepseekv3}}
    \label{fig:v3-model}
\end{figure}

Figure~\ref{fig:v3-model} presents the fundamental transformer block of DeepSeek-v3. The overall architecture comprises 61 layers, each incorporating two RMSNorm operations, a Multi-Head Latent Attention (MLA) block, and a linear layer. The linear layers exhibit a hybrid structure: the first three transformer layers utilize conventional feed-forward networks (FFN), while the remaining 58 layers implement MoE linear. 

\subsection{Structure Configuration}
Table~\ref{tab:structure_config} provides an overview of DeepSeek-v3's structural configuration. Our analysis focuses on memory consumption during training using FP16/BF16 formats, as FP8 training implementation remains outside the current scope. Consequently, certain FP8-related parameters, such as quantization scaling factors, are not included in this analysis.

\begin{table}[!h]
\captionsetup{skip=5pt} % 设置表格与标题之间的间距
\caption{Structure configuration of DeepSeek-v3}
\centering
\small
\renewcommand{\arraystretch}{1.2} % 调整行间距为1.5倍
\begin{tabular}{l|l|l|r}
\toprule
\textbf{Notation} & \textbf{Representation} & \textbf{Configuration of DeepSeek-v3}  & \textbf{Value} \\ \hline
$h$ & hidden dimension & $hidden\_size$ & 7168     \\
$h_E$ & hidden dimension of MoE's MLP & $moe\_intermediate\_size$ & 2048 \\
$h_F$ & hidden dimension of non-MoE's MLP & $intermediate\_size$ & 18432 \\
$d_h$ & dimension per head & $qk\_nope\_head\_dim$ & 128 \\
$n_h$ & No. of attention heads & $num\_attention\_heads$ & 128 \\
$d_{cq}$ & query compression dimension & $q\_lora\_rank$ & 1536 \\
$d_{hr}$ & per-head dimension of q/k for rope & $qk\_rope\_head\_dim$ & 64 \\
$d_c$ & key-value compression dimension & $kv\_lora\_rank$ & 512 \\
$N$ & No. of routed experts in MoE layer & $n\_routed\_experts$ & 256 \\
$N_s$ & No. of shared experts in MoE layer & $n\_shared\_experts$ & 1 \\
$l$ & No. of transformer layers & $num\_hidden\_layers$ & 61 \\
$v$ & vocabulary size & $vocab\_size$ & 129280 \\
\bottomrule
\end{tabular}
\label{tab:structure_config}
\end{table}

Table~\ref{tab:param-matrix} delineates the dimensional specifications of parameter matrices within the MLA and MoE components. Our analysis concentrates on the MoE layers' configurations, as these constitute the focal point of our memory consumption study. The first three transformer layers, which employ standard FFN, are significantly smaller in size compared to their MoE counterparts and are therefore excluded from this analysis.

\begin{table}[!h]
\captionsetup{skip=5pt} % 设置表格与标题之间的间距
\caption{Shape of parameter matrices of MoE transformer block}
\centering
\small
\renewcommand{\arraystretch}{1.2} % 调整行间距为1.5倍
\begin{tabular}{l|l|l|r}
\toprule
\textbf{Components} & \textbf{Parameter Matrix} & \textbf{Shape}  & \textbf{Of DeepSeek-v3} \\ \hline
\multirow{2}{*}{MLA} 
& $W^{DQ}$   & $[d_{cq}, \ h]$             & $[1536,\ 7168]$   \\
& $W^{UQ}$   & $[d_h * n_h, \  d_{cq}]$   & $[16384,\ 1536]$  \\
& $W^{QR}$   & $[d_{hr} * n_h,\  d_{cq}]$  & $[8192,\ 1536]$   \\
& $W^{DKV}$  & $[d_c,\  h]$              & $[512,\ 7168]$    \\
& $W^{UK}$   & $[d_h * n_h,\  d_c]$   & $[16384,\ 512]$   \\
& $W^{KR}$   & $[d_{hr},\  h]$             & $[64,\ 7168]$     \\
& $W^{UV}$   & $[d_h * n_h,\  d_c]$    & $[16384,\ 512]$   \\
& $W^O$    & $[h,\  d_h * n_h]$      & $[7168,\ 16384]$  \\ \hline
\multirow{2}{*}{MoE} 
& $gate\_proj$   & $[h, \ h_E]$             & $[7168,\ 2048]$   \\
& $up\_proj$   & $[h, \ h_E]$   & $[7168,\ 2048]$   \\
& $down\_proj$   & $[h_E, \ h]$  & $[2048,\ 7168]$    \\
\bottomrule
\end{tabular}
\label{tab:param-matrix}
\end{table}

\section{Model Parameter Counting}

\subsection{Layer-level Counting}
Table~\ref{tab:couting-layer} presents a detailed quantitative analysis of each component's parameters per layer, expressing their memory footprint in both megabytes (MB) and gigabytes (GB) under FP16/BF16 precision. The aggregate parameter count (187,107,328) and corresponding memory requirements for the MLA components are derived directly from the dimensional specifications presented in Table~\ref{tab:param-matrix}. Unlike some transformer architectures, DeepSeek-v3's training maintains separate parameters for the input token embedding matrix in the first layer and the output projection matrix in the final layer, i.e., word embeddings are not tied.

\begin{table}[!h]
\caption{Model parameter counting at layer-level (dtype: BF/FP16)}
\centering
\renewcommand{\arraystretch}{1.2} % 调整行间距为1.5倍
\small
\begin{tabular}{lllllll}
\toprule
\textbf{Layers} & \textbf{Modules} & \textbf{Shapes} & \textbf{No. Parameters} & \textbf{Per Layer} & \textbf{MB} & \textbf{GB} \\
\midrule
Layer 0 & Embedding & [129280, 7168] & 926,679,040  & 1.5 B & 2880 & 2.8 \\
& MLA & - & 187,107,328 & & & \\
& MLP & 3 * [7168, 18432] & 396,361,728 & & & \\
& LN & 2*7168+1536+512 & 16,384 & & & \\
\midrule
Layers 1 - 2 & MLA & - & 187,107,328  & 0.58 B & 1112 & 1.1 \\
& MLP & 3 * [7168, 18432] & 396,361,728 & & & \\
& LN & 2*7168+1536+512 & 16,384 & & & \\
\midrule
Layers 3 - 59 & MLA & - & 187,107,328 & 11.5 B & 21950 & 21.44 \\
& Gate & [256, 7168] & 1,835,008 & & & \\
& MoE & 3 * [7168, 2048] * 257 & 11,318,329,344 & & & \\
& LN & 2*7168+1536+512 & 16,384 & & & \\
\midrule
Layer 60 & MLA & - & 187,107,328 & 12.4 B & 23712 & 23.16 \\
& Gate & [256, 7168] & 1,835,008 & & & \\
& MoE & 3 * [7168, 2048] * 257 & 11,318,329,344 & & & \\
& LN & 2*7168+1536+512 & 16,384 & & & \\
& Head & [7168, 129280] & 926,679,040 & & & \\
\midrule
Total & & & & 671 B & 1,280,000 & 1250 \\
\bottomrule
\end{tabular}
\label{tab:couting-layer}
\end{table}

\subsection{Stages of Pipeline Parallel}
To quantify the peak memory utilization per GPU device, we use \texttt{PP16} pipeline parallel as our case study, which is the same as DeepSeek's official configuration. We focus on identifying and analyzing the pipeline stage that encompasses the maximum parameter volume. As detailed in Table~\ref{tab:pp-stage}, Stages 1-14 exhibit identical architecture and collectively constitute the largest parameter footprint, with each stage requiring 86 GB for static parameter storage.

A detailed examination of any single stage within Stages 1-14  provides sufficient information to determine the global peak memory requirements per GPU. For example, Stage 1 contains four sequential layers (layers 4-7). Each layer implements a MoE architecture with the following consecutive components: \texttt{RMSNorm}, \texttt{MLA}, \texttt{RMSNorm}, and \texttt{FFN-MoE} (MoE-enhanced feed-forward network).

\begin{table}[!h]
\captionsetup{skip=5pt} % 设置表格与标题之间的间距
\caption{Per-stage memory demands of model parameters under PP16 (dtype: BF/FP16)}
\centering
\small
\renewcommand{\arraystretch}{1.2} % 调整行间距为1.5倍
\begin{tabular}{l|l|l|r}
\toprule
\textbf{Stage}  & \textbf{No. Layers Per Stage} & \textbf{No. Params Per Stage}  & \textbf{Size in GB} \\ \hline
Stage $0$ & 4 & 14.16 B & 26  \\
Stages $1-14$ & 4 & 46 B & 86 \\
Stage 15 & 1 & 12.4 B & 23 \\
Sum & 61 & 671 B & 1250 \\
\bottomrule
\end{tabular}
\label{tab:pp-stage}
\end{table}

\section{Static Parameters Analysis}
Note that all the following analysis focuses on a single \texttt{PP} Stage. Firstly, we examine the device-level static parameter partitioning, which is primarily governed by three parallelization strategies: Tensor Parallelism (\texttt{TP}), Expert Parallelism (\texttt{EP}), and Expert Tensor Parallelism (\texttt{ETP}). For this investigation, we utilize the parallel configuration outlined in Table~\ref{tab:parallel-policy} as a reference case. The implementation of \texttt{TP} follows the method established in Megatron-LM \cite{narayanan2021efficient}, which defines the transformer block incorporating MLA and MoE components as follows.
\clearpage
\begin{python}
# megatron-lm/megatron/core/models/gpt/gpt_layer_specs.py:#L94
if multi_latent_attention:
    return ModuleSpec(
        module=TransformerLayer,
        submodules=TransformerLayerSubmodules(
            input_layernorm=TENorm,
            self_attention=ModuleSpec(
                module=MLASelfAttention,
                params={"attn_mask_type": AttnMaskType.causal},
                submodules=MLASelfAttentionSubmodules(
                    linear_q_proj=TEColumnParallelLinear,
                    linear_q_down_proj=TENoParallelLinear,
                    linear_q_up_proj=(
                        TELayerNormColumnParallelLinear
                        if qk_layernorm
                        else TEColumnParallelLinear
                    ),
                    linear_kv_down_proj=TENoParallelLinear,
                    linear_kv_up_proj=(
                        TELayerNormColumnParallelLinear
                        if qk_layernorm
                        else TEColumnParallelLinear
                    ),
                    core_attention=TEDotProductAttention,
                    linear_proj=TERowParallelLinear,
                    q_layernorm=IdentityOp,
                    kv_layernorm=IdentityOp,
                ),
            ),
            self_attn_bda=get_bias_dropout_add,
            pre_mlp_layernorm=TENorm if num_experts else IdentityOp,
            mlp=mlp,
            mlp_bda=get_bias_dropout_add,
        ),
    )
\end{python}

\begin{table}[!h]
\captionsetup{skip=5pt} % 设置表格与标题之间的间距
\caption{Parallel configuration used in case study}
\centering
\small
\renewcommand{\arraystretch}{1.2} % 调整行间距为1.5倍
\begin{tabular}{l|l|r}
\toprule
\textbf{Notation}   & \textbf{Short For} & \textbf{Value} \\ \hline
\texttt{DP}	& data parallelism	& 32 \\
\texttt{TP}	& tensor parallelism &	2 \\
\texttt{PP}	& pipeline parallelism & 16 \\
\texttt{EP}	& expert parallelism & 8 \\
\texttt{ETP}	& expert tensor parallelism & 1 \\
\texttt{EDP}	& expert data parallelism & 8 \\
\bottomrule
\end{tabular}
\label{tab:parallel-policy}
\end{table}

\subsection{RMSNorm $1/2$}
In the \texttt{PP16@TP2} parallel configuration, both \texttt{RMSNorm} components within each layer are replicated across all \texttt{TP} ranks, operating independently of \texttt{TP} partitioning, i.e., each \texttt{TP} rank maintains complete parameters for 8 \texttt{RMSNorm} operations distributed across 4 consecutive layers.

Therefore, the total parameters per rank can be calculated as follows:
\begin{itemize}
    \item per-layer parameter composition: $(7168 \times 2 + 1536 + 512) = 16,\!384 $
    \item total parameters across 4 layers: $(7168 \times 2 + 1536 + 512) \times 4 = 65,\!536$
\end{itemize}

Using BF16 (or FP16) format, which requires 2 bytes per parameter, the total memory footprint per GPU for \texttt{RMSNorm} parameters amounts to $65,\!536 \times 2 = 131,\!072$ bytes.

\subsection{MLA}
The implementation of MLA in Megatron-LM exhibits subtle architectural variations from the standard MLA employed in DeepSeek-v3. While these differences influence our analysis, their impact on overall memory estimation remains manageable for several reasons: (1) The parameter volume ratio between MLA and MoE components is notably disproportionate (approximately $1.8 : 113$);
(2) Consequently, any estimation discrepancies in the MLA component have small impact on the global memory analysis.

According to the Megatron-LM implementation of MLA, we partition the following parameters using \texttt{TP}. (1) \texttt{$W^{UQ}$}, \texttt{$W^{UK}$}, \texttt{$W^{UV}$}, and \texttt{$W^{O}$} are split across TP ranks. (2) \texttt{$W^{DQ}$}, \texttt{$W^{DKV}$}, \texttt{$W^{QR}$}, and \texttt{$W^{KR}$} are replicated on each rank without TP partitioning. Under the \texttt{PP16@TP2} configuration, the parameter storage requirements per \texttt{TP} rank are calculated as follows:
\begin{itemize}
    \item \textbf{TP Partitioned Parameters}: $(16384 \times 1536 + 16384 \times 512 \times 2 + 7168 \times 16384) \times 4 / 2 = 318,\!767,\!104$
    \item \textbf{Replicated Parameters}: $(1536 \times 7168 + 512 \times 7168 + 8192 \times 1536 + 64 \times 7168) \times 4 = 110,\!886,\!912$
    \item \textbf{Total Parameters}: $318,\!767,\!104 + 110,\!886,\!912 = 429,\!654,\!016$
\end{itemize}

Using the \texttt{BF16} data type, the memory requirement per GPU is: $429,\!654,\!016 \times 2 = 859,\!308,\!032 \text{ bytes.}$

\subsection{FFN-MoE}
The MoE layer consists of two components: the Router and the Experts. The Router parameters are not partitioned using TP, with a total parameter count of $256 \times 7168 = 1,\!835,\!008$. For the MoE component, under the \texttt{PP16@EP8@ETP1} configuration: Each stage contains 4 layers. The 256 routing experts per layer are evenly distributed across 8 ranks, resulting in 32 routing experts per rank. The shared expert is replicated across all ranks (as seen in bellow code). With \texttt{ETP1} the parameter matrices of individual experts are not partitioned using \texttt{TP}.

\begin{python}
    # megatron-lm-bc/megatron/core/transformer/moe/moe_layer.py: #L112
    if self.use_shared_expert:
        self.shared_experts = build_module(self.submodules.shared_experts, config=self.config)
\end{python}

Thus, each \texttt{EP} rank stores:
\[
4 \text{ layers} \times (32 \text{ routing experts} + 1 \text{ shared expert}) = 132 \text{ experts.}
\]
The total parameter count per rank for the experts is:
\[
132 \times 3 \times 7168 \times 2048 = 5,\!813,\!305,\!344.
\]
Combining the Router and Experts components, the total parameter count per rank is:
\[
1,\!835,\!008 \times 4 + 5,\!813,\!305,\!344 = 5,\!820,\!645,\!376.
\]
The memory requirement per GPU is: $ 5,\!820,\!645,\!376 \times 2 = 11,\!641,\!290,\!752 \text{ bytes.} $

\subsection{Model Parameters Per Device}
Table~\ref{tab:model-params-summary} summarizes the memory consumption of static parameters per device.

\begin{table}[!h]
\captionsetup{skip=5pt} % 设置表格与标题之间的间距
\caption{Model Parameters Per Device: Summary (dtype: BF/FP16)}
\centering
\small
\renewcommand{\arraystretch}{1.2} % 调整行间距为1.5倍
\begin{tabular}{l|l|l|l|l|r}
\toprule
\textbf{Modules}  & \textbf{No. Params Per Device} & \textbf{Bytes Per Device}  & \textbf{KB} & \textbf{MB} & \textbf{GB} \\ \hline
\texttt{RMSNorm} $1\&2$ & 65,536 & 131,072 & 128 & - & - \\
\texttt{MLA} & 429,654,016 & 859,308,032	& -	& 819.5	& - \\
\texttt{Non-MoE Part} &	429,719,552 &	859,439,104 &	- &	819.7 &	- \\
\texttt{MoE} & 5,820,645,376 & 11,641,290,752 & - & 11,102 & 10.84 \\
\midrule
\textbf{Total} &	6,250,364,928 &	12,500,729,856 & - & \textbf{11,922} & \textbf{11.64} \\
\bottomrule
\end{tabular}
\label{tab:model-params-summary}
\end{table}

\section{DeepSpeed ZeRO}
DeepSpeed ZeRO \cite{rasley2020deepspeed, rajbhandari2020zero} currently supports three strategies: \texttt{os}, \texttt{os+g}, and \texttt{os+g+params}. \textbf{\texttt{os}} shards the optimizer states across the \texttt{DP} groups; \textbf{\texttt{os+g}} shards both the optimizer states and gradients across the \texttt{DP} groups; \textbf{\texttt{os+g+params}} shards the optimizer states, gradients, and model weights across the \texttt{DP} groups.

The memory consumption per device depends on which of these strategies is employed. It is important to note that the \texttt{DP} and \texttt{EDP} configurations can differ, and thus the memory requirements of MoE part and non-MoE part must be calculated separately under ZeRO optimizations. The following analysis of the ZeRO strategies is based on the parallel configuration described in Table~\ref{tab:parallel-policy}. Table~\ref{tab:dtype} shows the data type used in the case study.

\begin{table}[!h]
\captionsetup{skip=5pt} % 设置表格与标题之间的间距
\caption{Data type used in the case study}
\centering
\renewcommand{\arraystretch}{1.2} % 调整行间距为1.5倍
\begin{tabular}{lllr}
\toprule
\textbf{Data} &  & \textbf{Type} & \textbf{Bytes Per Param/Value} \\ \hline
Weights & & BF16 & 2 \\
Activation & & BF16 & 2 \\
Gradients & &	FP32 & 4 \\ 
\midrule
Optimizer &	- Copy of parameters & FP32 & 4 \\ 
 &	- Momentum & BF16 & 2 \\
 &	- Variance & BF16 &	2 \\
\bottomrule
\end{tabular}
\label{tab:dtype}
\end{table}

Table~\ref{tab:zero-analysis} illustrates the memory consumption per device under different ZeRO optimization strategies, where the baseline model without ZeRO requires 11.64 GB for parameters, 23.3 GB ($6,\!250,\!364,\!928 \times 4$) for optimizer states, and 46.6 GB ($6,\!250,\!364,\!928 \times 8$) for gradients. The implementation of various ZeRO strategies progressively reduces memory usage. The \texttt{os} strategy reduces the memory consumption of optimizer states per device to 5.52 GB, as 
\[
(\frac{429,\!719,\!552}{32\ \texttt{DP}} + \frac{5,\!820,\!645,\!376}{8\ \texttt{EDP}}) \times 8.
\]

The \texttt{os+g} further reduces gradients memory per device to 2.76 GB, as 
\[
(\frac{429,\!719,\!552}{32\ \texttt{DP}} + \frac{5,\!820,\!645,\!376}{8\ \texttt{EDP}}) \times 4. 
\]

Finally, the \texttt{os+g+params} further reduces parameter memory per device to 1.38 GB, as
\[
(\frac{429,\!719,\!552}{32\ \texttt{DP}} + \frac{5,\!820,\!645,\!376}{8\ \texttt{EDP}}) \times 2.
\]

\begin{table}[!h]
\captionsetup{skip=5pt} % 设置表格与标题之间的间距
\caption{Memory consumption with different ZeRO optimizations}
\centering
\small
\renewcommand{\arraystretch}{1.2} % 调整行间距为1.5倍
\begin{tabular}{l|l|l|l|r}
\toprule
\textbf{ZeRO}   & \textbf{Static Parameters} & \textbf{Gradients} & \textbf{Optimizer} & \textbf{P+G+O} \\ \hline
\texttt{None}	& 11.64 GB & 23.3 GB & 46.6GB &	81.54 GB \\
\texttt{os}	& 11.64 GB	& 23.3 GB & 5.52 GB	& 40.46 GB \\
\texttt{os+g}	& 11.64 GB	&  2.76 GB & 5.52 GB	& 19.92 GB \\
\texttt{os+g+params}	& 1.38 GB & 2.76 GB	& 5.52 GB & 9.66 GB \\
\bottomrule
\end{tabular}
\label{tab:zero-analysis}
\end{table}

\section{Activation Analysis}
The activation memory should account for the recomputation strategy \cite{korthikanti2023reducing}. Recomputation introduces a little bit complexity; for instance, in selective recomputation, it is necessary to determine how many layers to recompute, which specific layers to target, and which components within each layer should be recomputed. Table~\ref{tab:config_4_activation} shows configurations used in the case study of activation memory analysis. In this report, we consider the two native cases: 
\begin{itemize}
    \item \textbf{No Recomputation}: No activation data is recomputed, and all intermediate activations are stored in memory.
    \item \textbf{Full Recomputation}: All intermediate activations are recomputed during the backward pass, minimizing memory usage at the cost of additional computational overhead.
\end{itemize}

\begin{table}[!t]
\captionsetup{skip=5pt} % 设置表格与标题之间的间距
\caption{Configurations of activation analysis}
\centering
\renewcommand{\arraystretch}{1.2} % 调整行间距为1.5倍
\begin{tabular}{l|l|l|r}
\toprule
\textbf{Notation}                    & \textbf{Representation} & \textbf{Of DeepSeek-v3}  & \textbf{Value} \\ \hline
$b$ & micro batch size & - & 1/2/4     \\
$s$ & sequence length & - & 4096 \\
$N_r$ & number of routed experts for each token & $n\_routed\_experts$ & 8 \\
$N$ & number of experts in each MoE layer & $num\_experts$ & 256 \\
$E_{token}$ & avg. No. of tokens processed by a single expert & - & $bs*N_r/N$ \\
\texttt{SP} & sequence parallelism  & - & On, 2 \\
\texttt{CP} & context parallelism  & - & 1 \\
\texttt{AC} & activation recomputation & - & None, Full \\
\bottomrule
\end{tabular}
\label{tab:config_4_activation}
\end{table}

\subsection{MLA}
\begin{figure}[!h]
    \centering
    \includegraphics[width=0.99\linewidth]{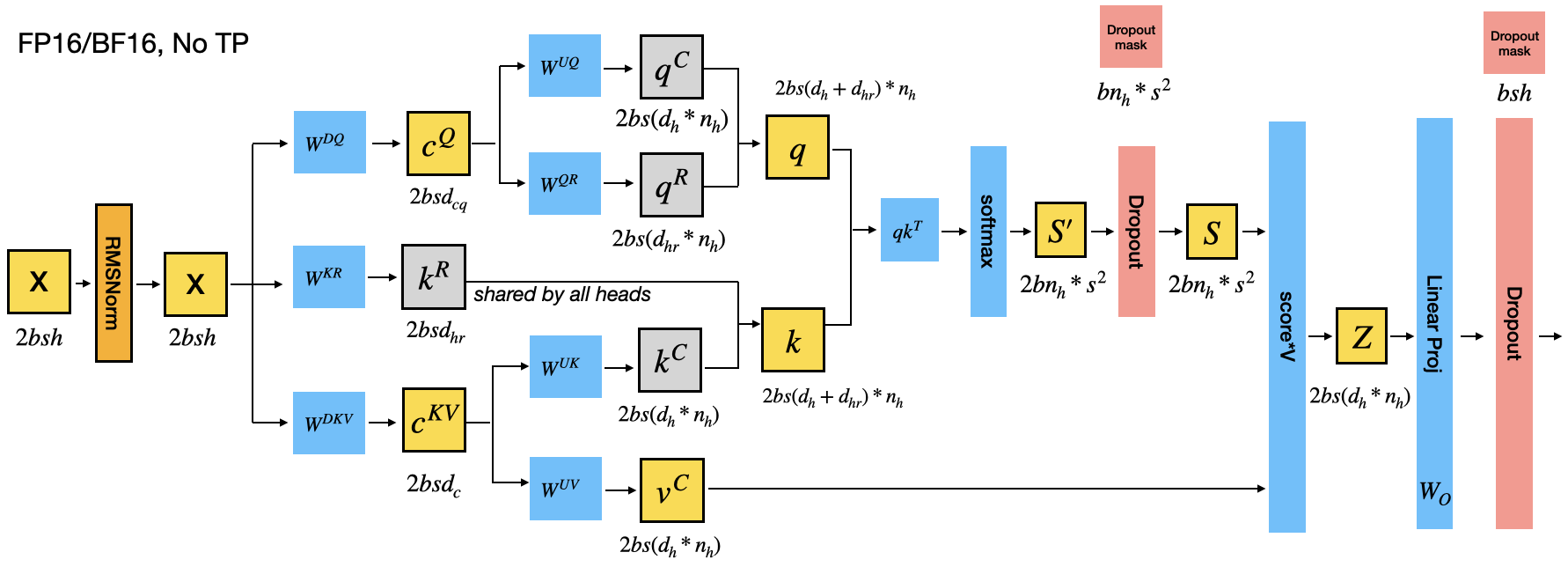}
    \caption{Activation pattern of MLA}
    \label{fig:mla_activation}
\end{figure}

Figure~\ref{fig:mla_activation} illustrates the activation patterns during MLA computation, where the total activation size (in bytes) without any parallelization is given by the following formula:

\[
4bsh + 2bs(d_{cq} + d_c) + 4bs(d_h + d_{hr}) \cdot n_h + 2bs(d_h \cdot n_h) + 5bn_hs^2 + 2bs(d_h \cdot n_h) + bsh
\]

The activation memory of MLA component under the parallel strategy of \texttt{TP2@SP2@CP1} reveals significant differences between non-recomputation and full-recomputation. Without recomputation, the total activation size per GPU for a single layer is:
\[
M^A_1=\frac{4bsh}{2}+2bs(d_{cq}+d_c)+\frac{4bs(d_h+d_{hr})n_h}{2} + \frac{2bs(d_hn_h)}{2} + \frac{5bn_hs^2}{2} +\frac{2bs(d_hn_h)}{2}+\frac{bsh}{2}
\]

Notably, the term $2bs(d_{cq}+d_c)$ remains undivided by \texttt{SP} due to the replication of weights $W^{DQ}$, $W^{DKV}$, $W^{QR}$, and $W^{KR}$ across ranks. For a four-layer \texttt{PP} stage, the total memory required per device for MLA activation becomes 
\[
4M^A_1=10bsh+8bs(d_{cq}+d_c)+16bsd_h*n_h+8bsd_{hr}n_h + 10bn_hs^2.
\]

In contrast, with the full-recomputation, the memory footprint reduces significantly to $M^A_2=2bsh/2$ per layer, as only the initial inputs before RMSNorm for both Attention and MLP components are retained, resulting in $4M^A_2=4bsh$ for four layers.

\subsection{MoE Linear}
\begin{figure}[!h]
    \centering
    \includegraphics[width=0.99\linewidth]{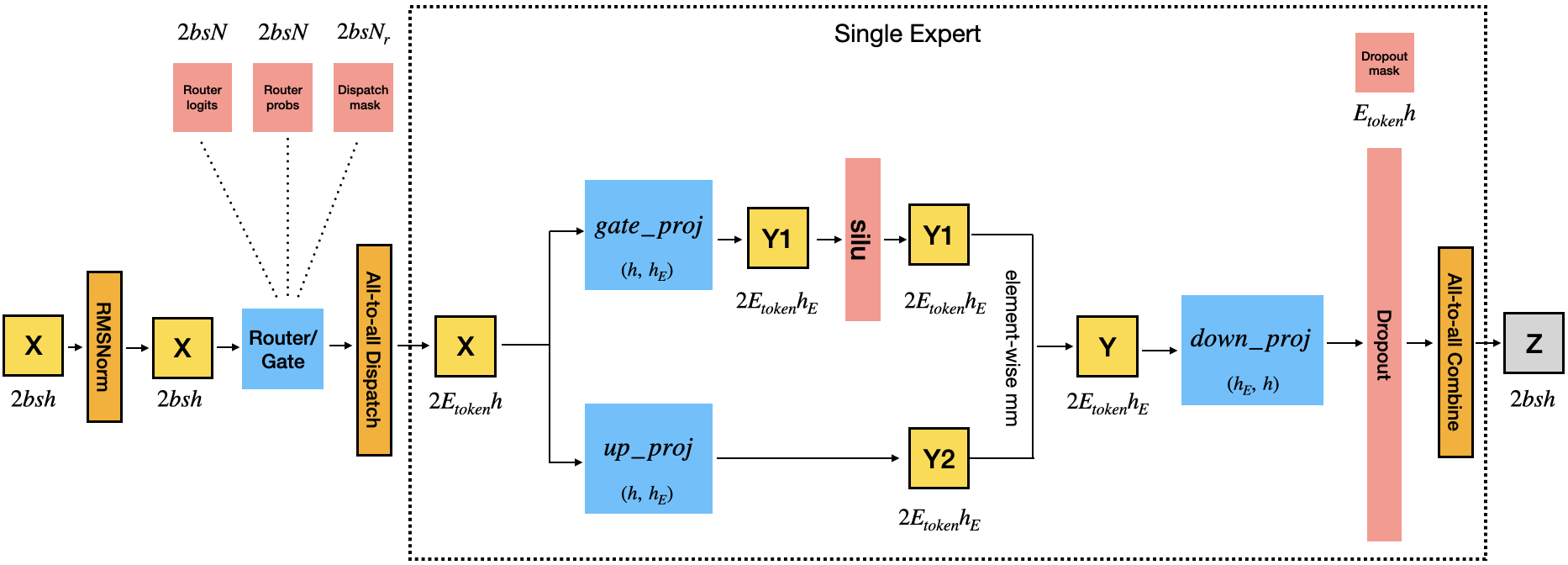}
    \caption{Activation pattern of MoE linear}
    \label{fig:moe_activation}
\end{figure}

For MoE layers with balanced load distribution, the average number of tokens processed by a single expert per MoE layer and micro batch is estimated by
\[
E_{token} = \frac{bs*N_r}{N}.
\]
Figure~\ref{fig:moe_activation} shows the activation pattern of MoE linear. The activation memory of MoE linear per layer under \texttt{SP2@EP8@ETP1} configuration without recomputation is:
\[
M^E_1=4bsh/2 + 4bsN + 2bsN_r + 32*(3*E_{token}*h + 8*E_{token}*h_E) + 1(3bsh+8bsh_E), 
\]
which simplifies to
\[
M^E_1 = 5bsh + 4bsN + 2bsN_r + bs*N_r/N *(96h+256h_E) + 8bsh_E
\]
after substituting $E_{token}$. For a four-layer \texttt{PP} stage, the total activation memory per device amounts to 
\[
4M^E_1=20bsh + 16bsN + 8bsN_r + 4bs*N_r/N *(96h+256h_E) + 32bsh_E.
\]

% the total activation size per GPU for a single MoE layer is $M_1 = 5bsh + 4bsN + 2bsN_r + bs*N_r/N *(96h+256h_E) + 8bsh_E$, where $b$ is batch size, $s$ is sequence length, $h$ is hidden dimension, $N$ is the total number of experts, $N_r$ is the number of routed experts, and $h_E$ is the expert hidden dimension. This equation accounts for both routed experts ($E_{token} = bs*N_r/N$) and shared experts ($E_{token} = bs$). 

With full recomputation, the memory footprint reduces to $M^E_2=bsh + 2bsN_r$ per layer, maintaining the Router outputs for consistency. The four-layer configurations require $4M^E_2=4bsh + 8bsN_r$.

\subsection{Activation memory per device}
Table~\ref{tab:activation_per_device} summarizes the above analysis for activation.

\begin{table}[!h]
\captionsetup{skip=5pt} % 设置表格与标题之间的间距
\caption{Activation memory per device}
\centering
\small
\renewcommand{\arraystretch}{1.2} % 调整行间距为1.5倍
\begin{tabular}{l|l|r}
\toprule
\textbf{Components}   & \textbf{AC None} & \textbf{AC Full} \\ \hline
\texttt{MLA}	      & $10bsh+8bs(d_{cq}+d_c)+16bsd_h*n_h+8bsd_{hr}n_h + 10bn_hs^2$	& $4bsh$ \\
\texttt{MoE}	& $20bsh + 16bsN + 8bsN_r + 4bs*N_r/N *(96h+256h_E) + 32bsh_E$ &	$4bsh + 8bsN_r$ \\
\midrule
Total & $4(M^A_1+M^E_1)$ & $8bsh + 8bsN_r$ \\
\bottomrule
\end{tabular}
\label{tab:activation_per_device}
\end{table}

\section{Temporal Buffer and Fragmentation}
Training memory estimation in practical implementations is affected by two additional factors. First, memory fragmentation, which typically ranges from 5\% to 30\% of the total allocated memory, introduces overhead that must be considered in training analysis. Second, temporary communication buffers required for inter-GPU data transfer generally occupy between 0.8 GB to 2 GB per device, depending on the specific parallel configuration and communication patterns. 

%it is prudent to include a buffer factor of at least 1.3x on the calculated memory requirements to account for these real-world constraints.

\clearpage
% Once you're familiar with the editor, you can find various project setting in the Overleaf menu, accessed via the button in the very top left of the editor. To view tutorials, user guides, and further documentation, please visit our \href{https://www.overleaf.com/learn}{help library}, or head to our plans page to

\bibliographystyle{unsrt}
\bibliography{reference}

\begin{thebibliography}{1}

\bibitem{liu2024deepseekv3}
Aixin Liu, Bei Feng, Bing Xue, Bingxuan Wang, Bochao Wu, Chengda Lu, Chenggang
  Zhao, Chengqi Deng, Chenyu Zhang, Chong Ruan, et~al.
\newblock Deepseek-v3 technical report.
\newblock {\em arXiv preprint arXiv:2412.19437}, 2024.

\bibitem{liu2024deepseekv2}
Aixin Liu, Bei Feng, Bin Wang, Bingxuan Wang, Bo~Liu, Chenggang Zhao, Chengqi
  Dengr, Chong Ruan, Damai Dai, Daya Guo, et~al.
\newblock Deepseek-v2: A strong, economical, and efficient mixture-of-experts
  language model.
\newblock {\em arXiv preprint arXiv:2405.04434}, 2024.

\bibitem{narayanan2021efficient}
Deepak Narayanan, Mohammad Shoeybi, Jared Casper, Patrick LeGresley, Mostofa
  Patwary, Vijay Korthikanti, Dmitri Vainbrand, Prethvi Kashinkunti, Julie
  Bernauer, Bryan Catanzaro, et~al.
\newblock Efficient large-scale language model training on gpu clusters using
  megatron-lm.
\newblock In {\em Proceedings of the International Conference for High
  Performance Computing, Networking, Storage and Analysis}, pages 1--15, 2021.

\bibitem{rasley2020deepspeed}
Jeff Rasley, Samyam Rajbhandari, Olatunji Ruwase, and Yuxiong He.
\newblock Deepspeed: System optimizations enable training deep learning models
  with over 100 billion parameters.
\newblock In {\em Proceedings of the 26th ACM SIGKDD International Conference
  on Knowledge Discovery \& Data Mining}, pages 3505--3506, 2020.

\bibitem{rajbhandari2020zero}
Samyam Rajbhandari, Jeff Rasley, Olatunji Ruwase, and Yuxiong He.
\newblock Zero: Memory optimizations toward training trillion parameter models.
\newblock In {\em SC20: International Conference for High Performance
  Computing, Networking, Storage and Analysis}, pages 1--16. IEEE, 2020.

\bibitem{korthikanti2023reducing}
Vijay~Anand Korthikanti, Jared Casper, Sangkug Lym, Lawrence McAfee, Michael
  Andersch, Mohammad Shoeybi, and Bryan Catanzaro.
\newblock Reducing activation recomputation in large transformer models.
\newblock {\em Proceedings of Machine Learning and Systems}, 5:341--353, 2023.

\end{thebibliography}

\end{document}